\documentclass[12pt]{article}

\usepackage{setspace}
\usepackage{amsmath, amssymb, amsthm,  float,graphicx}
\def\K{{\mathcal K}}

\def\be{\begin{equation}}
\def\ee{\end{equation}}
\def\lp{\ell_P}
\def\R{{\mathcal R}}
\def\R{{\mathcal R}}

\def\L{{\mathcal L}}

\def\L{{\mathcal L}}
\def\K{{\mathcal K}}
\def\be{\begin{equation}}
\def\ee{\end{equation}}
\def\lp{\ell_P}

\def\l{\lambda}
\def\l{\lambda}

\def\beq{\begin{eqnarray}}\def\eeq{\end{eqnarray}}

\begin{document}
\title{\bf Entanglement entropy in higher derivative holography}

\author{ Arpan Bhattacharyya,  Apratim Kaviraj and  Aninda Sinha\\
\it Centre for High Energy Physics,
\it Indian Institute of Science,\\ \it C.V. Raman Avenue, Bangalore 560012, India. \\
}

\maketitle
\vskip 1cm
\begin{abstract}{\small We consider holographic entanglement entropy in higher
derivative gravity theories. Recently Lewkowycz and Maldacena \cite{maldacena} have
provided a method to derive the equations for the entangling surface from
first principles. We use this method  to compute the entangling surface in
four derivative gravity. Certain interesting differences compared to the two
derivative case are pointed out.  For Gauss-Bonnet gravity, we show that in the regime where this
method is applicable, the resulting equations coincide with proposals in
the literature as well as with what follows from considerations of the
stress tensor on the entangling surface. Finally we demonstrate that the
area functional in Gauss-Bonnet holography arises as a counterterm needed
to make the Euclidean action free of power law divergences.}
\end{abstract}

\tableofcontents
\onehalfspace
\section{Introduction}
In the context of holography, Ryu and Takayanagi \cite{ryu, rev} proposed a remarkably simple prescription to compute entanglement entropy (EE) for the field theory dual to Einstein gravity. One computes the minimal area for the entangling surface which extends into the bulk from an area functional \footnote{A covariant version was proposed in \cite{mukund}.}. When the corresponding area functional is evaluated on a black hole horizon it gives the black hole entropy. If the entangling surface is spherical, there exists an independent derivation of holographic entanglement entropy due to Casini, Huerta and Myers \cite{chm}. This derivation holds quite generally since it maps the calculation of the entanglement entropy of spherical regions to the Wald entropy of hyperbolic black holes \cite{myersme}. Unfortunately this generalization is not known for more general entangling surfaces.

 In field theory one frequently uses the replica trick \cite{calabrese} namely,
\be
S_{EE}=\lim_{n\rightarrow 1} S_n=\lim_{n\rightarrow 1} \frac{1}{1-n} \log {\rm tr}~\rho^n\,,
\ee
where $\rho$ is the reduced density matrix. The $S_n$'s are the Renyi entropies. From the holographic side, this entails being able to do calculations for general $n$ and then doing an analytic continuation to $n=1$. Fursaev suggested a derivation \cite{fursaev} of the Ryu-Takayanagi prescription which appeared to have some problems since it led to the wrong Renyi entropies \cite{head} (there is a proposed resolution to this in \cite{fur1}). Until recently, it was not clear if the implementation of this on the gravity side was consistent as it appeared to give rise to singularities in the bulk leading to inconsistencies (see \cite{bum1, bum2} for a solution to this problem in AdS$_3$). 

Unlike black hole entropy where for a general theory of gravity there exists  the Wald formula \cite{wald}, there is no such elegant generalization for holographic entanglement entropy. The only case where there are arguments to extend the Ryu-Takayanagi minimization prescription is that of Lovelock theories \cite{higher,higher2}. It is not known for instance if for an arbitrary theory of gravity the minimization prescription holds. It is thus of interest and importance to extend the holographic entanglement entropy prescription to general higher derivative theories of gravity.

Recently, Lewkowycz and Maldacena \cite{maldacena} have proposed a way to set up the calculation of entanglement entropy in what appears to be a self-consistent manner. They introduced the concept of a generalized gravitational entropy which is an extension of the usual Euclidean methods applicable to black hole entropy calculation to solutions where there is no $U(1)$ symmetry. They observed that when one considers $n=1+\epsilon$ with $\epsilon\rightarrow 0$ in the gravity solutions, quite generally there would be singularities in the equations of motion which can be shown to vanish if the minimal area conditition is satisfied. We  ask what happens if their procedure is extended to higher derivative theories. We will consider four derivative theories of gravity. In particular we will consider Gauss-Bonnet gravity as this is the simplest higher derivative theory with nice properties. We will also consider more general four derivative theories and show that the above method only works for Gauss-Bonnet gravity .

Furthermore, recently it was observed \cite{AB} in the context of Einstein gravity that if the Brown-York tensor \cite{stress} is evaluated on the co-dimension one entangling surface (not setting $t=0$), then setting the time-time component of this tensor to zero gives  identical equations for the surface as what follows from the Ryu-Takayanagi prescription. The simple reason for this is that for static surfaces, the co-dimension one extrinsic curvature trivially decomposes: $^{(d)}K_a^a= ^{(d)}\!\!\!K_t^t+ ^{(d-1)}\!\!K_i^i$. Thus $^{(d)}\!K_t^t-h_t^t \,{}^{(d)}\!K_a^a=0$ leads to $^{(d-1)}\!K_i^i=0$ which is the same as the minimal surface condition for the $d-1$ slice used in the Ryu-Takayanagi calculation. But the combination $^{(d)}\!K_t^t-h_t^t \,{}^{(d)}\!K_a^a$ is nothing but the $tt$ component of the usual Brown-York  tensor evaluated on the co-dimension one slice in $d$-dimensions. This equivalence is of course certainly not guaranteed for higher derivative theories. Thus it is interesting to ask what happens in the Gauss-Bonnet theory where the Gibbons-Hawking surface term is known \cite{kcube}.

The primary questions of interest in this paper are:

\begin{enumerate}
\item In higher derivative theories, does the equation for the entangling surface derived from the generalized gravitational entropy prescription agree with existing proposals?
\item What role does the area functional play in Euclidean gravity calculations?
\item For general four derivative theories, is there an extension to the Ryu-Takayanagi area functional?
\end{enumerate}

The paper is organized as follows. In section 2, we summarize the proposals for entanglement entropy in Gauss-Bonnet gravity. We also show what the equation for the entangling surface is if we set $T_{tt}^{BY}=0$ on the co-dimension one entangling surface. In section 3 we extend the derivation of the minimal area due to Lewkowycz and Maldacena to Gauss-Bonnet gravity. We also consider more general four derivative gravity. In section 4, we show that the area functional in Gauss-Bonnet gravity is a suitable counterterm to remove power law divergences in an Euclidean gravity calculation. We conclude in section 5. There are several appendices with more calculational details.

\section{Entanglement entropy proposals in GB gravity}
 The five dimensional bulk action (Euclidean) supplemented  by the appropriate surface and counterterms is given by:
\be \label{totact}
I_{tot}=I_{bulk}+I_{GH}+I_{ct}\,.
\ee
\be \label{bulk}
I_{bulk}= -\frac{1}{2\lp^{3}} \int d^{5}x \, \sqrt{g}\big[\hat R+\frac{12}{L^{2}} + \lambda \frac{L^{2}}{2}GB\big]\,.\\
\ee 
 $\lambda$ is the Gauss-Bonnet coupling, $GB=\hat R_{\alpha \beta \mu\nu}\hat R^{\alpha \beta \mu\nu}-4 \hat R_{\alpha\beta}\hat R^{\alpha\beta}+\hat R^{2}$ and $g$ is the determinant of the bulk metric.\footnote{ Here hat denotes 5 dimensional quantity}
Surface terms are needed  for imposing Dirichlet boundary condition on the metric. $I_{GH}$ is the usual surface term for Gauss-Bonnet gravity \cite{kcube}
\be\label{surface1}
I_{GH}=- \frac{1}{\lp^{3}} \int d^{4}x \sqrt{\gamma}\big[K-\lambda L^{2}(2G_{ab}K^{ab}+\frac{1}{3}(K^{3}-3KK_{2}+2K_{3})\big]\,.\\
\ee
Here $K_{2}=K_{ab}K^{ab}$ and $K_{3}=K^{a}_{b}K^{b}_{c}K^{c}_{a}$. $\gamma$ is the determinant of the four dimensional boundary slice.  $K_{ab}=e^{\alpha}_{a}e^{\delta}_{b} p^{\beta}_{\delta}\nabla_{\alpha} n_{\beta}$ is the extrinsic curvature, $K=K_a^a$,  $p^{\alpha}_{\gamma}=g^{\alpha}_{\gamma}-n^{\alpha}n_{\gamma}$ is the projection operator, $n_{\beta}$ is the normal for the surface\,. We will use Greek letters for the bulk indices. For a four dimensional slice we will use $a,b,c,d$  and   $i,j,k,l$  for the indices used for a three dimensional slice.

\be \label{metric}
ds^{2}=\frac{\tilde L^{2}}{ z^{2}}(dz^{2}+dt^{2}+h_{ij}dx^{i}dx^{j})
\ee
where, $\tilde L$ is  the $AdS$ radius and $h_{ij}$ is a three dimensional metric given below. For the calculation of EE  for a spherical entangling surface we write the boundary $h_{ij}$ in spherical polar coordinates as,
\be
^{sphere}h_{ij}dx^{i}dx^{j}= d\rho^{2}+\rho^{2}d\Omega_{2}^{2}\,,
\ee 
where $d\Omega_{2}^2=d\theta^{2}+\sin^2 \theta d\phi^{2}$ is the metric of a unit two-sphere and
$\theta \in [0,\pi]$ and $\phi\in [0, 2\pi]\,.$\\
For a cylindrical entangling surface,
\be
^{cylinder}h_{ij}dx^{i}dx^{j}=du^{2}+d\rho^{2}+\rho^{2}d\phi^{2}\,.
\ee$u$ is coordinate along the direction of the length of the cylinder. For a cylinder of length $H$, $u\in [0,H]\,.$ 
Here $\tilde L=\frac{L}{\sqrt{f_{\infty}}}\,,$ where $f_{\infty}$ satisfies the following equation $1-f_{\infty}+f_{\infty}^{2}\lambda=0\,.$ We will take the smallest positive root which is continuously connected to $f_{\infty}=1$ for the Einstein ($\lambda =0$) case.

The counterterm action  $I_{ct}$ is needed for the cancellation of the power law divergences in $I_{tot}$ when the boundary is $z=\epsilon$. For our case this works out to be \cite{yale,bhss}
\be
I_{ct}=\frac{1}{\lp^{3}} \int d^{4}x\, \sqrt{\gamma}\big[ c_1\frac{3}{\tilde L}+ c_2 \frac{\tilde L}{4}R\big]\,,
\ee
where $R$ is the  four dimensional Ricci scalar and $ c_1=1-\frac{2}{3}f_{\infty}\lambda $ and $c_2=1+2f_{\infty}\lambda\,.$
There are two trace anomaly coefficients given by \cite{gbholo}
$$ a= \frac{\pi^{2}L^{3}}{f_{\infty}^{3/2}\lp^{3}}(1-6f_{\infty}\lambda)\,,  \quad c=\frac{\pi^{2}L^{3}}{f_{\infty}^{3/2}\lp^{3}}(1-2f_{\infty}\lambda) \,.$$

If the entangling surface is a sphere then the universal part of the holographic entanglement entropy is proportional to $a$ while if the surface is a cylinder then it is proportional to $c$ \cite{solod}. This can be taken to be tests that any proposal for entanglement entropy in four dimensions should satisfy.


\subsection{Modified area functional $S_{JM}$}
For Gauss-Bonnet gravity, it was argued that the following area functional\footnote{Jacobson and Myers \cite{jm} had obtained this functional in the context of black hole entropy from a Hamiltonian approach. For black holes, this gave the same result as Wald entropy although it differs by terms dependent on the square of the extrinsic curvature. It was shown in \cite{higher} that the area functional arising from Wald entropy would not give the correct universal behaviour for entanglement entropy. In particular for cylindrical surfaces it would pick out $a$ rather than $c$ as the proportionality constant in the universal term.} would give rise to the correct universal parts of the entanglement entropy for spherical and cylindrical entangling surfaces\cite{higher, higher2}:
\be \label{JM}
S_{JM}= \frac{2\pi}{\ell_{p}^{3}}\int d^3 x \sqrt{\Sigma}\big[1+\lambda L^{2}\,\mathcal{R}\big]\,,\\
\ee
where $\mathcal{R}$  is  the scalar curvature defined for the $3$ dimensional surface obtained by putting $\rho=f(z)$ and $ t=0$ in eq.(\ref{metric}) to obtain a co-dimension two surface. $\Sigma$ is the determinant of the metric of this three dimensional slice. One has to extremize $S_{JM}$ which will determine $f(z)$. 

\subsubsection*{General minimal surface condition}
Let us begin with a general analysis  independent of the actual form of the entangling surface to derive the minimal surface condition from eq.(\ref{JM}) along the lines of \cite{mukund}\,. 
For  the co-dimension two surface we have the pullback metric $$\Sigma_{ij}= g_{\mu\nu}\frac{\partial X^{\mu}}{\partial x^{i}}\frac{\partial X^{\nu}}{\partial x^{j}}\,,$$ where $X^{\mu}$ and $x^{i}$ are respectively bulk and the boundary coordinates. Next we consider a general action $S=\int d^{3}x\L$ constructed out of $\Sigma_{ij}$.  In order to extremize this, we vary $S$ with respect to $\Sigma_{ij}$ where the variation of $\Sigma_{ij}$ comes from the variation of $X^{\mu}$.
\begin{align}
\begin{split}
\delta S=\int d^{3}x \frac{\delta\L}{\delta \Sigma_{ij}}\delta \Sigma_{ij}&=\int d^{3}x\frac{\delta\L}{\delta \Sigma_{ij}}(g_{\mu\nu}\partial_{i}\delta X^{\mu}\partial_{j}X^{\nu} +g_{\mu\nu}\partial_{i}X^{\mu}\partial_{j}\delta X^{\nu})\\
&=\int d^{3}x \big[-2\partial_{i}(  \frac{\delta\L}{\delta \Sigma_{ij}} \partial_{j} X^{\nu}g_{\mu\nu})\delta X^{\mu}+\rm{total\,derivative}\big]\,.
\end{split}
\end{align}
We  drop the surface term assuming that an appropriate Gibbons-Hawking term exists.  At this stage, following \cite{mukund}, we choose the following coordinate system $$ X^{\mu}=(x^{i},y^{l})$$ where $y^{l}$ denotes the two  transverse directions to the surface. We have,
\begin{align}
\begin{split}
\Pi_{\mu}=\frac{\delta S}{\delta X^{\mu}}&=- 2 \partial_{i}(\frac{\delta\L}{\delta \Sigma_{ij}} g_{j\mu})=-2\partial_{i}(\frac{\delta\L}{\delta \Sigma_{ij}})g_{j\mu}-2\frac{\delta\L}{\delta \Sigma_{ij}}\partial_{i}g_{j\mu}\,.
\end{split}
\end{align}
 Using  $ \nabla_{i} \frac{\delta\L}{\delta \Sigma_{ij}}=0$ and $\nabla_{i}g_{j\mu}=0\,,$ we get
\begin{align}
\begin{split}
\Pi_{\mu}&=2\Gamma^{i}_{ik}\frac{\delta\L}{\delta \Sigma_{kj}}g_{j\mu}-2\Gamma^{k}_{i\mu}g_{jk}\frac{\delta\L}{\delta \Sigma_{ij}}
\end{split}
\end{align}
Contracting with the normal\footnote{ $\mu$ labels the component and $m$ labels the transverse directions.} $\,^{(m)}n^{\mu}$  ,
\begin{align}
\begin{split}
^{(m)}n_{\mu}\Pi^{\mu}=-2 \,^{(m)}n^{\mu}\Gamma^{k}_{i\mu}g_{jk}\frac{\delta\L}{\delta \Sigma_{ij}}=0\,.
\end{split}
\end{align}
In this coordinate system $^{(m)}n^{\mu}\Gamma^{k}_{i\mu}=-\K_i^k$ where $\K_{ij}$ is the 3 dimensional extrinsic curvature. So when $S=S_{JM}$ we get the following condition for the minimal surface, 
\be \label{condition1}
\K+\lambda L^{2}(\R\K-2\R_{ij}\K^{ij})=0\,.
\ee
We have also checked this by explicitly computing the surface equations for cylinder, sphere and slab geometries.

\subsubsection*{EE for a sphere}We will use the  metric in (\ref{metric}) with $ ^{sphere}h_{ij}$ as $h_{ij}$. The equation of motion for the entangling surface can be found either using eq.(\ref{condition1}) or by directly extremizing $S_{JM}$. This yields
\begin{align}
\begin{split} \label{myerssphere}
&-(1+f'(z)^2)\big[6 \lambda f_{\infty} z^2 f'(z)+3 f(z)^2 f'(z) (1-2 f_{\infty} \lambda +f'(z)^2)\\&+2 z f(z) (1-2 f_{\infty} \lambda +(1+4 f_{\infty} \lambda ) f'(z)^2)\big]+z\big [6f_{\infty}\lambda  z\big( z +2   f(z) f'(z)\big)\\&+f(z)^2 (1-2 f_{\infty} \lambda +(1+4 f_{\infty} \lambda ) f'(z)^2)\big] f''(z)=0\,.
\end{split}
\end{align}
The solution of (\ref{myerssphere}) is, \be f(z) =\sqrt{z_0^{2}-z^{2}}\,,\ee where $z_0$ is the radius of the entangling surface. The surface closes off when $z=z_h=z_0$. Then evaluating $S_{JM}$ one gets the universal $\log$ term as,
\be
S_{EE}=- 4 a \ln(\frac{z_{0}}{\epsilon})\,.\\
\ee

\subsubsection*{EE for the cylinder}
 Starting with the metric in (\ref{metric})  and with $^{cylinder}h_{ij}$ as $h_{ij}$ and proceeding similarly as before we can determine $f(z)$\,.
The equation satisfied by $f(z)$ is given by
\begin{align}
\begin{split} \label{myerscyl}
&-(1+f'(z)^2)\big[z(1-2 f_{\infty}  \lambda )+f'(z)(z (1+4 f_{\infty} \lambda ) f'(z)+3 f(z) [1-2 f_{\infty} \lambda +f'(z)^2])\big]\\&+z [6 f_{\infty} \lambda z   f'(z)+f(z)(1-2 f_{\infty} \lambda +(1+4 f_{\infty} \lambda ) f'(z)^2)] f''(z)=0\,.
\end{split}
\end{align}
The solution of (\ref{myerscyl}) is,  \be f(z)= z_{0}\big(1-\frac{z^{2}}{4 z_{0}^{2}}+f_{4}z^{4}+ \frac{1}{32 z_{0}^{4}} z^{4}\ln z\big)+\cdots\ee 
where  $f_{4}$ can be determined numerically.
Then evaluating $S_{JM}$ one gets the universal $\log$ term for entanglement entropy as,
\be
S_{EE}=- \frac{c}{2}\frac{H}{z_{0}} \ln(\frac{z_{0}}{\epsilon})\,.\\
\ee
In this case the surface closes off when $z=z_h\neq z_0$.

\subsection{Stress tensor considerations}
As explained above \cite{AB}, the minimal surface condition in Einstein gravity is identical to setting the $tt$ component of the Brown-York tensor on the co-dimension one $\rho=f(z)$ surface to zero.  A heuristic explanation for this finding is given in fig.1. If we want an analogy with MERA \cite{swingle}, then we need to note that at each level in MERA we are in the ground state. $T_{tt}=0$ on the co-dimension one slice is a condition that will guarantee this \footnote{This condition is reminiscent of the way that Einstein equations emerge from thermodynamical considerations \cite{jacob}.}. 
\begin{figure}[htpb]
\centering\includegraphics[height=2.2in,width=7in ]{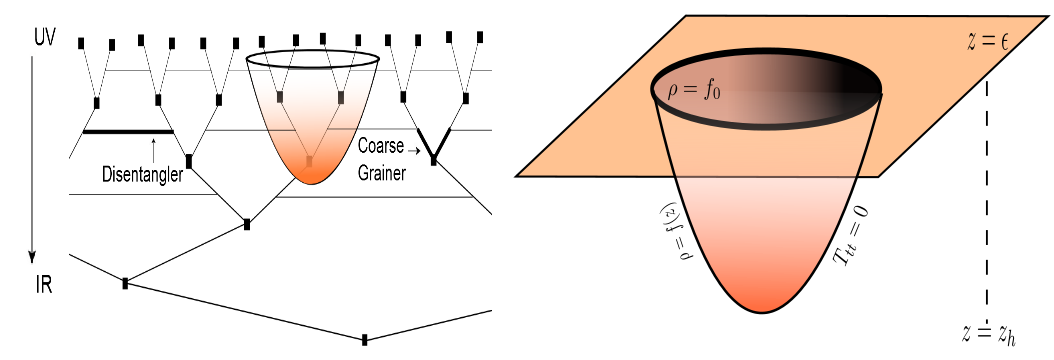}
\caption{$T_{tt}=0$ determines the shape which also implies the ground state of the configuration. This has a similarity with the MERA procedure in which all the states in a particular level are in the ground state.}\label{mera_tensor}
\end{figure}

 We wish to ask if setting $T_{tt}=0$ in the Gauss-Bonnet case gives rise to the same equations or not. The stress tensor expression is given by \cite{ Liu},
\be \label{sstress}
T^{BY}_{ab}=\frac{1}{\ell_{p}^{3}}[K_{ab}-\gamma_{ab}K+\lambda L^{2} (q_{ab}-\frac{1}{3}\gamma_{ab}q)]\,,
\ee
where $$ q=h^{ab}q_{ab}\,, \, \, q_{ab}=2 K K_{ac}K^{c}_{b} -2K_{ac} K^{cd}K_{db}+K_{ab}(K_{2}-K^{2})+2K R_{ab}+RK_{ab}-2K^{cd}R_{cadb}-4R_{(a}^{c}K_{b)c}\,. $$
We will  set $T_{tt}=0$ and see what the equations are. 

\subsubsection*{Entangling surface condition}
 In order to compare with the covariant expression in eq.(\ref{condition1}), we need to do a 3+1 split as shown in appendix A. After this $3+1$ splitting, we get quantities only made up of $3$- extrinsic and $3$-intrinsic curvatures. The resulting condition is given by
\be \label{condition2}
 \K+\lambda L^{2}\big[(\R\K-2 \R_{ij}\K^{ij})+\frac{1}{3}(-\K^{3}+3 \K \K_{2}-2\K_{3})\big]=0\,.\\
\ee
Here $\K_2=\K_a^b \K_b^a$ and $\K_3=\K_a^b \K_b^c \K_c^a$. If the extrinsic curvature is small compared to the intrinsic curvatures, then the $O(\K^3)$ terms can be ignored and we will get precisely eq.(\ref{condition1}). 


\subsubsection*{Sphere}
For the spherical entangling surface we find that $T_{tt}=0$ leads to
\begin{align}
\begin{split} \label{ttsphere}
0= & (1+f'(z)^2) \big[4 f_{\infty} \lambda z^2  f'(z)+f(z)^2 f'(z) (3 (1-2 \lambda f_\infty )\\&+(3-2f_{\infty}\lambda) f'(z)^2\big]+2 z f(z) \big[(1-2f_{\infty}\lambda) +(1+2f_{\infty}\lambda) f'(z)^2)\big]\\&-z\big[4 \lambda f_\infty  z\big(z  +2 f(z) f'(z)\big)+f(z)^2 \big((1-2f_{\infty}\lambda)\\&+(1+2 f_{\infty}\lambda) f'(z)^2\big)\big] f''(z)\,.\\
\end{split}
\end{align}
$f(z)=\sqrt{z_{0}^2-z^2}$ is an exact solution to the above equation. However the equations (\ref{ttsphere}) and (\ref{myerssphere}) are not the same. The difference is given by
\be
2 f_{\infty} \lambda   \left(z+f(z) f'(z)\right) \left(f'(z)+f'(z)^3-z f''(z)\right)\,.
\ee 
This extra contribution comes precisely due to the $\K^{3}$ terms in $T_{tt}$. For a general surface, we have $\K\propto 1/z_{0}$. So in the large $z_{0}$ limit, the $O(\K^3)$ terms are subleading compared to the $O( \K)$ ones. The $O(\K^3)$ difference in the equations is to be noted since we will find a similar difference using the generalized gravitational entropy method as well.

\subsubsection*{Cylinder}
For the cylinder we find
\begin{align}
\begin{split} \label{ttcyl}
0= & (1+f'(z)^2)\big [z-2  f_{\infty} \lambda z  +f'(z) \big(z\, (1+2 f_{\infty}\lambda) f'(z)+f(z)[3(1-2f_{\infty}\lambda) \\&+(3-2 f_{\infty}\lambda) f'(z)^2]\big)\big]-z\big [4 f_{\infty} \lambda z   f'(z)+f(z) \big((1-2f_{\infty}\lambda) \\&+(1+2 f_{\infty}\lambda)  f'(z)^2\big)\big] f''(z)\,.
\end{split}
\end{align}
Solving (\ref{ttcyl}) we get,
\be
f(z)=z_{0}\big(1-\frac{1}{4 z_{0}^{2}} z{}^2+f_{4} z^4+\frac{1}{32 z_{0}^{4}} z^4\ln z\big)+\cdots\,,
\ee
where as before $f_{4}$ can be determined numerically. This is the same solution as what arises from eq.(\ref{myerscyl}). However the equations for the surface are not the same. The difference  between eq.(\ref{ttcyl}) and eq.(\ref{myerscyl}) is
\be \label{diff}
2 f_{\infty} \lambda  f'(z) \left(z+f(z) f'(z)\right) \left(f'(z)+f'(z)^3-z f''(z)\right)\,.
\ee
 Again if we throw away the $\K^{3}$ term assuming a large $z_{0}$, we get the same equation  eq.(\ref{myerscyl}). Here we must emphasize the fact that the solution coming from $T_{tt}=0$ method in this case  agrees with that of coming from $S_{JM}$ upto $z^4 \ln z$ term around $z=0$  but not in the neighbourhood of $z=z_{h}$.


\section{Generalized gravitational entropy method}
 In a recent work, Lewkowycz and Maldacena \cite{maldacena} have shown that one can find the minimal surface condition by demanding that the $n=1+\epsilon$ solution satisfies the linearized equations of motion near the conical singularity.
In this section we will derive the  equation  for $f(z)$ for a general entangling surface for four derivative gravity following \cite{maldacena} . We start with the following five dimensional metric in the form used in Gaussian normal coordinates,
\be
ds^{2}=\tilde L^2(e^{2\rho} (dx_{1}^{2}+dx_{2}^{2})+g_{ij}dx^{i}dx^{j})= \tilde L^2(e^{2\rho} (dr^2+r^2 d\tau^2)+g_{ij}dx^{i}dx^{j})\,,
\ee
where 
$$g_{ij}=h_{ij}+x_{1} \,^{1}\K_{ij}+x_{2}\,^{2}\K_{ij}+ (x_1^2+x_2^2) Q_{ij}+\cdots.$$
 When $\rho=0$ this corresponds to a local solution of the equations of motion in the neighbourhood of $x_1=x_2=0$. 
In these coordinates $^{1,2}\K_{ij}$ are the extrinsic curvatures for the co-dimension two surface $x_1=x_2=0$. We have put in the $O(x^2)$ term as well since it will turn out to play a role in the general four derivative calculation. Both $\K_{ij}$ and $Q_{ij}$ can be expressed in terms of $g_{ij}$. Namely in radial coordinates $^{(r)}\K_{ij}=1/2 \partial_r g_{ij}|_{r=0}$ while $Q_{ij}=2 \partial^2_r g_{ij}|_{r=0}$.
The $\cdots$ will involve higher powers of $x_i$'s. To extract the leading singular behaviour these will not be needed. When $n=1+\epsilon$ we put $\rho=-\frac{\epsilon}{2} \log (x_1^2+x_2^2).$
Going to the complex coordinates,
$
w=\frac{x_1+ix_2}{2}, \bar{w}=\frac{x_1-ix_2}{2}
$
we have
$
g_{ij}=h_{ij}+w \,\,^{(w)}\K_{ij}+\bar w\,\,^{(\bar w)}\K_{ij}, \,^{(w)}\K=\displaystyle\frac{^{1}\K_{ij}+i \,\,^{2}\K_{ij}}{2},\,\,^{(\bar w)}\K =\displaystyle\frac{^{1}\K_{ij}-i\,\,^{2}\K_{ij}}{2}\,.
$
 For $\tau=0$, we have
$
^{(w)}\K_{ij}=\,\,^{(\bar w)}\K_{ij}\,.
$
Next we consider a general bulk lagrangian,
\be
I_{bulk}=\int d^{5}x \big [\hat R+\frac{12}{L^{2}}+ \frac{L^{2}}{2}(\lambda_{3}\hat R^{2}+\lambda_{2} \hat R_{\alpha \beta}\hat R^{\alpha\beta}+\lambda_{1}\hat R_{\alpha\beta\sigma\delta}\hat R^{\alpha\beta\sigma\delta})\big]\,.
\ee
For Gauss-Bonnet gravity, $\lambda_{1}=\lambda\,,\lambda_{2}=-4\lambda\,, \lambda_{3}=\lambda\,.$
The equations of motion (these can be found in many places, eg \cite{Sinha:2010pm}) are,
\be \label{eomfull}
G_{\alpha\beta}-\frac{6}{L^{2}}g_{\alpha\beta}-\frac{ L^{2}}{2}H_{\alpha\beta}=T_{\alpha\beta}
\ee
where,$$
G_{\alpha\beta}=\hat R_{\alpha\beta}-\frac{1}{2} g_{\alpha\beta} \hat R\,
$$
and
$$
 H_{\alpha\beta}=\lambda_3(-2 \hat R\hat  R_{\alpha\beta}+2 \hat\nabla_\alpha \hat\nabla_\beta \hat R +\frac{1}{2}g_{\alpha\beta}\hat R^2 - 2 g_{\alpha\beta} \hat\nabla^2 \hat R)
$$
$$+\lambda_2(\hat\nabla_{\alpha}\hat\nabla_{\beta}\hat R+2\hat R^{\delta \sigma}\hat R_{\delta \alpha \beta \sigma}  - \hat\nabla^2 \hat R_{\alpha\beta}+\frac{1}{2} g_{\alpha\beta}\hat  R_{\delta \sigma}\hat R^{\delta\sigma}-\frac{1}{2} g_{\alpha\beta} \hat\nabla^2 \hat R) 
$$
$$
+\lambda_1(\frac{1}{2}g_{\alpha\beta}\hat R_{\delta \sigma \mu 
\nu}\hat R^{\delta\sigma \mu\nu}-2\hat R_{\alpha \sigma \delta \mu}{\hat R_\beta}{}^{\sigma \delta 
\mu}-4\hat\nabla^2 \hat R_{\alpha \beta}+2\hat\nabla_\alpha \hat\nabla_\beta \hat R+4 \hat R_\alpha^\delta \hat R_{\beta \delta}+4\hat R^{\delta \sigma}\hat R_{\delta \alpha \beta \sigma})
$$
Here $\hat \nabla$ is the usual covariant derivative made of the full five dimensional metric. For the Gauss-Bonnet case, it is easy to see that all the $\hat\nabla \hat\nabla R$ terms cancel out.
In order to satisfy the linearized equations of motion, we generally have to consider metric perturbations $\delta g$ such that $\delta g(\tau)=\delta g(\tau+2\pi)\,.$  In the equations of motion, there will be divergences of the type $\epsilon/w$, $(\epsilon/w)^2$ etc as we will show explicitly below.  These divergences need to be canceled. Since the metric perturbation is periodic, we will need to argue that the contribution from the perturbations are zero \cite{maldacena}. Thus we will set to zero all the divergent terms. This will give us the equation for the surface. In the following all the curvature quantities are constructed using $h_{ij}$ and raising and lowering will be done through $h_{ij}$. 

 \par First we list the all possible divergences in the equations of motion for Gauss-Bonnet gravity.
\underline{Divergences in the $ww$ ($\bar w \bar w$) component} :
\begin{align}
\begin{split}
&-\frac{\epsilon}{2 w}\,^{(w)}\K-\frac{\lambda L^{2}\epsilon}{2 w}\big[\,^{(w)}\K\R-2 \,^{(w)}K_{ij}\R^{ij}+(-\,^{(w)}\K^3+3\,^{(w)}\K\,^{(w)}\K_2
-2\,^{(w)}\K_3)\big]\,.
\end{split}
\end{align}
Notice that if we wanted to set this to zero, the condition would be similar to what follows from $T_{tt}=0$ with precisely the same combination of the $O(\K^3)$ terms except for an overall factor of $1/3$. There are no $O(\K^3)$ terms from minimizing $S_{JM}$. This mismatch will be addressed very shortly.\\
\underline{Divergences in the $w\,i$ ($\bar w \, i$) component }:
\begin{align}
\begin{split}
&-\frac{\lambda L^{2}\epsilon}{2 w}\big[2\,^{(w)}\K\nabla_j(\,^{(w)}\K^j_i)-2\,^{(w)}\K \nabla_i (\,^{(w)}\K) +2\,^{(w)}\K^j_i\nabla_j (\,^{(w)}\K)\\
&-2\,^{(w)} \K_{ij}\nabla_k (\,^{(w)}\K^{kj})+2\,^{(w)} \K_{kj}\nabla_i (\,^{(w)}\K^{kj})-2\,^{(w)} \K_{jk}\nabla^j (\,^{(w)}\K^{k}_i)\big]\,.
\end{split}
\end{align}
\underline{Divergences in the $i\,j$ component} :
\begin{align}
\begin{split}
&2\lambda L^{2}\big[\frac{\epsilon}{w}(\,^{(w)}\K_{ij}\,^{(w)}\K_2-2\,^{(w)}\K_{ik}\,^{(w)}\K^{kl}\,^{(w)}\K_{lj}+\,^{(w)}\K_{il}\,^{(w)}\K^{l}_{j}\,^{(w)}\K-\,^{(w)}\K\,^{(w)}\K_2h_{ij}+\,^{(w)}\K_3h_{ij})\\&
+\frac{\epsilon^{2}}{w^2}(\,^{(w)}\K^2 h_{ij}-2\,^{(w)}\K\,^{(w)}\K_{ij}-\,^{(w)}\K_2h_{ij}+2\,^{(w)}\K_{ik}\,^{(w)}\K^{k}_{j})\big]\,.
\end{split}
\end{align}
$\R, \R_{ij}$ etc are made up of the metric $h_{ij}$\,.  The metric fluctuation will contribute  $-\frac{1}{2}(1-2f_{\infty}\lambda)\Box \delta g_{\alpha\beta}$  using transverse traceless gauge but this is zero using the periodicity argument as in \cite{maldacena}. Now to get the minimal surface condition we have to set all the divergences in the equation of motion to zero (the divergences in the $\bar w \bar w$, $\bar w i$ components are the same form as the $w w $ and $w i$ since we have set $\tau=0$).
Now notice the following:
\begin{itemize}
\item Unlike the Einstein-Hilbert case in \cite{maldacena}, in the Gauss-Bonnet case, there are divergences in the $ij$, $iw$ components as well. 
\item If we assume that $O(\K) \sim \alpha w/\epsilon$ where $\alpha\ll 1$, then the $ij$, $iw$ components go to zero.
\item In this case $O(\K^3)\ll O(\K)$ so the $\K^3$ terms can be dropped.
\item The $O(w\bar w)$ term proportional to $Q_{ij}$ does not enter the calculation.
\end{itemize}
Now setting the remaining $ww$ component of the equation of motion to zero we get,
\be
 \,^{(w)}\K+ \lambda L^{2}\big[\,^{(w)}\K\R-2 \R^{ij}\,^{(w)}K_{ij}\big]=0\,.
\ee
Thus for Gauss-Bonnet we get exactly the condition (\ref{condition1}) which follows from  $S_{JM}$! In fact since we are in a regime where $O(\K^3)\ll O(\K)$, this is also in agreement with what follows from $T_{tt}=0$ on the co-dimension one surface.

\vskip 1cm

Now we turn to the more general four derivative case. In this case, the contribution from the fluctuation of the metric is four derivative. In order to simplify our discussion, we will restrict our attention to the case where the four derivative terms are a combination of the Weyl-squared $C_{\alpha\beta\gamma\delta}C^{\alpha\beta\gamma\delta}$ and the Gauss-Bonnet. In this case the contribution from the fluctuation in the transverse traceless gauge can be shown \cite{deser}\footnote{The expansion around $r=0$ is obtained by considering $L\rightarrow \infty$.} to be proportional to $\Box^2 \delta g_{\alpha\beta}$. Then we can use the periodicity argument to set this to zero. To begin with, we will ignore the contribution from the second order terms in the metric proportional to $Q_{ij}$. Several problems in this general case will become apparent without introducing this. First let us list out the various divergences.

\noindent \underline{Divergences in the $ww$ ($\bar w\bar w$) component} :
\begin{align}
\begin{split}
&-\frac{\epsilon}{2 w}\,^{(w)}\K-\frac{L^{2}}{2}\big[\frac{\epsilon}{w}\big\{\lambda_{3}\,^{(w)}\K\R-(2\lambda_{1}+\lambda_{2}+4\lambda_{3}) \R^{ij}\,^{(w)}\K_{ij}-2(3\lambda_1+2\lambda_2+6\lambda_3)\,^{(w)}\K_3+(\lambda_2+7\lambda_3)\,^{(w)}\K\,^{(w)}\K_2\\&
-\l_3\,^{(w)}\K^3+(\frac{\lambda_{2}}{2}+2\lambda_{1})\nabla^{2}\,(^{(w)}\K)\big\}+\frac{\epsilon}{w^{2}}\big\{\,(2\lambda_1-\lambda_2-6\lambda_3)^{(w)}\K_2+\,(\lambda_2/2+2\lambda_3)^{(w)}\K^2\big\}\\&+\frac{\epsilon^{2}}{w^{2}}\big\{\,(-4\lambda_1+2\lambda_2+12\lambda_3)^{(w)}\K_2-\,(\lambda_2+4\lambda_3)^{(w)}\K^2\big\}+\frac{\epsilon^2 \log w}{w^2}\big\{\,2(2\lambda_1-\lambda_2-6\lambda_3)^{(w)}\K_2\\&+\,(\lambda_2+4\lambda_3)^{(w)}\K^2\big\}\big]\,.
\end{split}
\end{align}
\underline{Divergences in the $w \bar w$ component }:
\begin{align}
\begin{split}
&-\frac{L^{2}}{2}\big[\frac{\epsilon}{w}\big\{\,(2 \lambda_1-3\lambda_2-14\lambda_3)^{(w)}\K_2\,^{(w)}\K+\,4(-2\lambda_1+\lambda_2+6\lambda_3)^{(w)}\K_3+\frac{1}{2}(\lambda_{2}+4\lambda_{3})\,^{(w)}\K^{3} \big\}\\&+\frac{\epsilon^2}{w^2}\big\{(2\lambda_1-\lambda_2-6\lambda_3)\,^{(w)}\K_2+\,(\lambda_2/2+2\lambda_3)^{(w)}\K^2\big\}\big]\,.
\end{split}
\end{align}
Note that  it is evident from this expression that for the Gauss-Bonnet case, the divergences in the $w \bar w$ component vanishes.

\noindent\underline{Divergences in the $w\,i$ ($\bar w \, i$ ) component} :
\begin{align}
\begin{split}
&-\frac{L^2\epsilon}{2 w}\big[\,-2(5\lambda_1+3\lambda_2+6\lambda_3)^{(w)}\K_{kj}\nabla_i\,^{(w)}\K^{kj}+\,2(\lambda_1+\lambda_2+2\lambda_3)^{(w)}\K\nabla_i \,^{(w)}\K\\&
-\,(2\lambda_1+\lambda_2)^{(w)}\K_{ij}\nabla^j \,^{(w)}\K-2\lambda_1\,^{(w)}\K_{ij}\nabla_k\,^{(w)}\K^{kj}-\frac{\lambda_2}{2}\,^{(w)}\K\nabla_j\,^{(w)}\K^j_i-2\lambda_1\,^{(w)}\K_{jk}\nabla^j\,^{(w)}\K^k_i\big]\,.
\end{split}
\end{align}
\noindent \underline{Divergences in the $i,j$ component} : 
\begin{align}
\begin{split}
&-\frac{L^{2}}{2}\big[\frac{\epsilon}{w}{\big\{}\,4(-4\lambda_{1}+3\lambda_{3})^{(w)} \K_{2}\,^{(w)}\K_{ij}+\,4(\lambda_{1}-\lambda_{3})^{(w)}\K^{2}\,^{(w)}\K_{ij}-\,8(17\lambda_{1}+4\lambda_{2})^{(w)}\K_{il}\,^{(w)}\K^{l}_{j}\,^{(w)}\K\\&
+\,4(2\lambda_{2}+9\lambda_{1})^{(w)}\K_{ik}\,^{(w)}\K^{kl}\,^{(w)}\K_{lj}-4(11\lambda_{3}+3\lambda_{2})h_{ij}\,^{(w)}\K\,^{(w)}\K_{2}+4(24\lambda_{3}+6\lambda_{2}-\lambda_{1})h_{ij}\,^{(w)}\K_{3}\\&+(4\lambda_{3}+\lambda_{2})
h_{ij}\,^{(w)}\K^{3}\big{\}}+\frac{\epsilon^2}{w^2}\big{\{}\,8\lambda_{1}^{(w)}\K_{ij}^{(w)}\K-\,(40\lambda_{1}+8\lambda_{2})^{(w)}\K_{ik}\,^{(w)}\K^{k}_{j}+(3\lambda_{2}+8\lambda_{3})h_{ij}\,^{(w)}\K^{2}\\&+2(2\lambda_{1}-3\lambda_{2}-6\lambda_{3})h_{ij}\,^{(w)}\K_{2}\big{\}}\big]\,.\\
\end{split}
\end{align}

\noindent\underline{Effect of $\mathcal{O}(w\bar w)$ term in the metric}: \par\vspace{.3cm} As we noted earlier,  if we add a term of the form $w\bar w \,Q_{ij}$ to the metric $g_{ij}$ our Gauss-Bonnet results remain unaffected. For the general case the structures of the divergences due to $Q_{ij}$ in the equation of motion are shown schematically  below (the actual forms will not be needed in our discussion).\vspace{.3cm}

\noindent \underline{Divergences in the $ww$, $w\bar w$ and  $ij$ component}:
$$-\frac{L^{2}}{2}\big[\frac{\epsilon}{w} \,^{(w)}\K \,Q+\frac{\epsilon^{2}}{w^{2}}\,Q+\frac{\epsilon}{w^{2}}\,Q\big]\,. $$
\underline{Divergences in the $w\, i$ component}:
$$-\frac{L^{2}\epsilon}{2 w} \nabla_i Q\,.$$ 
Here we note the following:
\begin{itemize}
\item For the Weyl-squared combination we need to set $\lambda_2=-4 \lambda_1/3$ and $\lambda_3=\lambda_1/6$.

\item Unlike the Gauss-Bonnet case, there are $(O(\epsilon/w^2))$ terms in the $ww$ component. These arise due to $\nabla \nabla R$ terms. There is no such divergence in the other components except the one proportional to $Q_{ij}$. The logic in the previous case was to make an approximation such that the divergences in the other components of the equations vanish. Thus unless we make further assumptions, the
$O(\epsilon/w^2)$ terms will be the leading terms leading to wrong equations for the surface. If $O(\K)\sim \alpha w/\epsilon$ then the $O(\epsilon/w^2 \K^2)\sim \alpha^2/\epsilon$. So to be able to ignore these terms we need $\alpha^2/\epsilon\ll 1$.  Making this assumption already appears problematic.

\item To be able to drop the terms proportional to $Q_{ij}$ we will need to make yet further assumptions. If we demand $O(Q_{ij})\ll O( w^2/\epsilon)$, only then can the additional terms be dropped.

\item We can drop the rest of the divergences in the other components if we make the same weak extrinsic curvature assumption as in the Gauss-Bonnet case.

\end{itemize}

If we make all the above assumptions, then we can set the $ww$ component to zero to see what the equation for the surface is for the Weyl-squared case \footnote{In the previous versions, the coefficients of the $\R\K$ and $\R_{ab}\K^{ab}$ terms in eq.(35) had typos.}
\be
 ^{(w)}\K+\frac{\lambda_1}{6} L^{2}\big[{}^{(w)}\K\R-8^{(w)}\K_{ij}\R^{ij}+8\nabla^{2}(^{(w)}\K)\big]=0\,.
\ee
We have not been able to find an area functional leading to the above equation. It is unlikely that one exists. This is because if we construct one out of a function of $\R$ and/or $\K$, we will generally get terms that are proportional to  $\K$ (see section 2.1). Thus the $\nabla^2 \K$ term will be problematic. Only for the Gauss-Bonnet case is it possible to get an area functional  as in that case the $\nabla^{2}\K$ term vanishes and we get the combination $\R\K-2\R_{ij}\K^{ij}$\,. In this case the functional is $S_{JM}$ as shown in section 2.1. 


\section{ $S_{JM}$ as a counterterm}

In this section we will show that adding $S_{JM}$ as a counterterm in Gauss-Bonnet gravity and evaluating the Euclidean action by integrating outside a spherical entangling surface, the power law divergences will cancel out and the final result will be simply the universal part of the entanglement entropy. The analogous calculation for the cylinder case is given in the appendix. 
We will compute the onshell Euclidean action outside the entangling region. This will need us to add several contributions which we will list out one by one\footnote{In the context of EE in boundary conformal field theories \cite{bcft} a similar calculation is performed. However in that case the boundary conditions on the metric are Neumann.}. The total boundary in this example comprises of the $\rho=f(z)$ co-dimension one entangling surface and the $z=\epsilon$ surface which is the usual boundary where the field theory lives\footnote{There will also be contributions from the corners where the entangling surface meets the AdS boundary but these terms can be removed along the lines of \cite{bcft2}.}.
We
 start by evaluating the bulk action
around $z=0$ we get
\begin{eqnarray}I_{bulk}
&=& -\frac{4 \pi   \beta  L^3 (z_{0}^{3}-{ \Lambda}^{3}) (1-6 f_{\infty} \lambda )}{3 f_{\infty}^{3/2} \lp^3 \epsilon^4}+\frac{4 \pi  \beta  L^3 z_{0}( 1-6 f_{\infty} \lambda )}{f_{\infty}^{3/2} \lp^3 \epsilon^2}-\frac{2 \pi  \beta  L^3 (1-6 f_{\infty} \lambda )}{f_{\infty}^{3/2} \lp^3 \,z_{0}}\ln(\frac{z_{0}}{\epsilon})+\cdots \,. \nonumber\\
\end{eqnarray}
Here $\Lambda$ is a cut-off on the $\rho$ integral. It will cancel out in the end.
$I_{GH}$ is the surface term for $z=\epsilon$ slice leading to ,
\begin{align}
\begin{split}
I_{GH}
&=\frac{16 \pi  \beta  L^3  (z_{0}^{3}-{ \Lambda}^{3}) (1-2 f_{\infty} \lambda )}{3 f_{\infty}^{3/2} \lp^3 \epsilon^4}+\cdots\,,
\end{split}
\end{align}
$I_{ct}$ is the counterterm for $z=\epsilon$,
\begin{align}
\begin{split}
I_{ct}&=-\frac{4 \pi   \beta  L^3 (z_{0}^{3}-{\Lambda}^{3}) (3-2 f_{\infty} \lambda )}{3 f_{\infty}^{3/2} \lp^3 \epsilon^4}+\cdots\,.
\end{split}
\end{align}
$\tilde I_{GH}$ is the surface term  on the $\rho=f(z)$ slice.
Expanding $\tilde I_{GH}$ we get,
\begin{eqnarray}
\tilde I_{GH}
&=&-\frac{2\pi  \beta  L^3 z_{0} (1-6 f_{\infty} \lambda )}{\epsilon^2 f_{\infty}^{3/2} \lp^3}+\frac{2 \pi  \beta  L^3 (1-6 f_{\infty} \lambda) }{f_{\infty}^{3/2} \lp^3\, z_{0} }\ln(\frac{z_{0}}{\epsilon})+\cdots\,.\nonumber\\
\end{eqnarray}
 It is evident here that after performing the $z$ integral and identifying $ \beta=2\pi z_{0}$ , $\tilde I_{GH}$ has precisely the same structure as that of the entanglement entropy \cite{area} with an overall negative sign:
\be
\tilde I_{GH}=-\frac{4\pi^{2}   L^3 z_{0}^{2} (1-6 f_{\infty} \lambda )}{\epsilon^2 f_{\infty}^{3/2} \lp^3}+4\, a\ln(\frac{z_{0}}{\epsilon})+\cdots\,.
\ee   
At this stage if we add up the contributions, we find that the $O(1/\epsilon^4)$ terms have cancelled out as expected but we are still left with $O(1/\epsilon^2)$ terms. It is evident that to have a local counterterm with leading divergence $O(1/\epsilon^2)$ we will need a three dimensional one. Since $S_{JM}$ has a leading divergence $1/\epsilon^2$ we can add this piece to see what we get. 
To keep things a little general, let us add $$I_{JM}=\frac{\beta}{2\pi }\int d^{3}x \sqrt{\Sigma}[\alpha+\kappa \lambda  L^{2} \R]$$ evaluated on the co-dimension two surface $t=0, \rho=f(z)$, with the coefficients $\alpha$ and $\kappa$  arbitrary for now. Evaluating this we get,
\begin{align}
\begin{split}
I_{JM}
&=-\frac{4 \pi  \beta  L^3 z_{0} (\alpha-\kappa\, 6 f_{\infty} \lambda )}{f_{\infty}^{3/2} \lp^3 \epsilon ^2}+\frac{2 \pi   \beta  L^3 (\alpha-\kappa\, 6 f_{\infty} \lambda )}{f_{\infty}^{3/2} \lp^3 z_{0} }\ln(\frac{z_{0}}{\epsilon})+\cdots\,.
\end{split}
\end{align}
Adding them together, if we demand that all the power law divergences should cancel out we get  $\alpha=\kappa=1$\,. So we see that the counterterm  $I_{JM}$ needed to remove the power law divergences turns out to be $\frac{\beta}{2\pi}S_{JM}\,.$ Thus we get the total action to be
\be \label{tot}
I_{tot}= \frac{4\pi^2    L^3 (1-6 f_{\infty} \lambda ) }{ f_{\infty}^{3/2}  \lp^3 }\ln (\frac{z_{0}}{\epsilon} )+\cdots
\ee
If we use $T I_{tot}=E-T S_{EE}=-T S_{EE}$ in an analogy with black hole thermodynamics (and identify $T=1/(2\pi z_0)$ as the Unruh temperature), then $S_{EE}$ will be the universal piece of the entanglement entropy. Thus it appears that the area functional can be thought of as a suitable counterterm to remove power law divergences in Euclidean gravity calculation.

\section{Discussion}

In this paper we set out to examine holographic entanglement entropy in higher derivative gravity. We used the procedure recently proposed by Lewkowycz and Maldacena to derive the equations for the entangling surface. We compared the result obtained for a general four derivative theory of gravity with existing results in the literature. In doing so we find answers to the following:
\begin{enumerate}
\item {\bf  In higher derivative theories, does the equation for the entangling surface derived from the generalized gravitational entropy prescription agree with existing proposals?}

It turned out that the derivation for the entangling surface proposed by Lewkowycz and Maldacena works when the $O(\K^3)$ terms could be neglected compared to the $O( \K)$ terms. In this regime, the equations agreed with those in the literature. In particular they were in agreement with \cite{higher, higher2}. Further in this regime they also agreed with the $T_{tt}^{BY}=0$ condition \cite{AB}. One motivation for seeking such a condition is to see if the surface equations can arise from arguments which do not rely on the replica trick. Current methods rely crucially on the replica trick. 

\item {\bf What role does the area functional play in Euclidean gravity calculations?}

By considering the spherical entangling surface as an example, we found that the area functional regularizes the total Euclidean action of Gauss-Bonnet gravity by removing power law divergences in accordance with the findings in \cite{AB}. The action was evaluated outside the entangling surface in a naive analogy with the black hole entropy. Whether this was a coincidence or not needs further investigation (see also \cite{see1}). The main motivation in thinking that $S_{JM}$ could be a counterterm is the following observation: It is known from \cite{higher} that the difference between what arises from the area functional arising from Wald entropy and $S_{JM}$ is that the latter has no 3-extrinsic curvature terms. When we consider counterterms in AdS/CFT in a setup where we are imposing Dirichlet boundary conditions on the metric, there cannot be any extrinsic curvature terms in the counterterm action.

\item {\bf For general four derivative theories, is there an extension to the Ryu-Takayanagi area functional?}

This question of course is of great interest and relevance since the Wald formula for black holes is meant to work for an arbitrary theory with higher derivative corrections. Our investigations in this paper seem to suggest that the answer is no. At four derivative order, only for Gauss-Bonnet theory does there appear to be an area functional. This conclusion was reached by assuming the generalized gravitational entropy approach used in \cite{maldacena} to be valid. Also, we did not treat the higher derivative coupling perturbatively. In this case (which is of course more natural in the context of string theory), it may be possible to see the existence of an area functional perturbatively. For example, it is always possible in a perturbative framework, to perform a field redefinition and recast four derivative terms into Gauss-Bonnet (where there is an area functional).

Another point we should make is suppose we considered more general corrections of the form $V(\phi) R^2$. In this case, it appears that the form for $V(\phi)$ would affect the equations for the minimal surface. Thus the area functional, assuming one exists, cannot be purely geometrical---this of course could have been anticipated from the Wald formula for black hole entropy.

\end{enumerate}

There are several open questions. The most important one is to see if a modification of the Lewkowycz-Maldacena method can work for arbitrary extrinsic curvatures unlike what we found here where the extrinsic curvatures needed to be small. The implications for the (non) existence of an area functional for arbitrary theories of gravity also need to be investigated. Of course, entanglement entropy should satisfy strong subadditivity and it is not guaranteed that arbitrary theories of gravity with holographic duals (assuming they exist) should satisfy this property.  In type IIB string theory, there is a well known $R^4$ correction \cite{r4}. It will be interesting to extend the derivation to this case. In quasi-topological gravity \cite{quasi}, the radial equations are two derivative similar to Gauss-Bonnet. One can try to see if an area functional exists in this case. It will also be interesting to consider excited states \cite{takatemp} from a first principles approach. 

\vskip 1cm
{\bf Acknowledgments} : We thank Janet Hung,  Gautam Mandal  and Shiraz Minwalla for discussions. AS acknowledges support from a Ramanujan fellowship, Govt. of India.

\appendix
\section{Stress Tensor split}
In this section, we will present some details of the $3+1$ split of the $tt$ component of the stress tensor for the Gauss-Bonnet gravity. The $tt$ component of $T^{BY}$ can be split as,
\begin{align}
\begin{split}
T_{tt}&=\frac{1}{\lp^{3}}\big(K_{tt}-\gamma_{tt}K-\lambda L^{2}\big[\gamma_{tt}(RK-2R_{ab}K^{ab})-RK_{tt}-2R_{tt} K+4R^{c}_{(t}K_{t)c}+2R_{tatb}K^{ab}\\&+\frac{1}{3}\gamma_{tt}(-K^{3}+3KK_{2}-2K_{3})+ K^{2}K_{tt}-2KK^{c}_{t}K_{tc}-K_{tt}K_{2}+2K_{t}^{a}K_{a}^{b}K_{b t}\big]\big)\,.
\end{split}
\end{align}
Then we split all the quantities along time and the rest.
\begin{align}
\begin{split}
R_{tt}&=R_{tit}{}^{i}\,, R_{ij}=R_{itj}{}^{t}+\R_{ikj}{}^{k}\,,\\
R&=R_{ab} \gamma^{ab}=2R^{t}_{t}+\R^{i}_{j}\,.\\
\end{split}
\end{align}
Similarly,
\begin{align}
\begin{split}
K=K^{t}_{t}+\K^{i}_{i}\,.\\
\end{split}
\end{align}
Using this relations in $T_{t}^{t}\,,$
\begin{align}
\begin{split}
K_{t}^{t}-\gamma^{t}_{t}K&=K^{t}_{t}-\gamma^{t}_{t}(K^{t}_{t}+\K^{i}_{i})=-\gamma^{t}_{t}K^{i}_{i}\,.
\end{split}
\end{align}
Now,
\begin{align}
\begin{split}
  &\gamma_{t}^{t}(RK-2R_{ab}K^{ab})-RK_{t}^{t}-2R_{t}^{t}K+4R^{c}_{t}K_{c}^{t}+2R^{t}{}_{atb}K^{ab}\\&=\gamma_{t}^{t}[(2R^{t}_{t}+\R^{i}_{i})(K^{t}_{t}+\K^{i}_{i})-2R_{tt}K^{tt}-2R^{t}{}_{itj}\K^{ij}]-2\R_{ij}\K^{ij}\\&-(2R^{t}_{t}+\R^{i}_{i})K^{t}_{t}-2R^{t}_{t}(K^{t}_{t}+\K^{i}_{i})+4 K^{t}_{t}R^{t}_{t}+2R^{t}{}_{itj}\K^{ij}\\
&=\R\,\K-2 \R_{ij}\K^{ij}\,.
\end{split}
\end{align}
Here our metric is diagonal. $\R^{}$ and $\K$ are the scalar and extrinsic curvature on the co-dimension two slice.
Also after decomposition, 
\be
K^{2}K^{t}_{t}-2KK^{ct}K_{tc}-K^{t}_{t}K_{2}+2K^{t}_{ a}K^{ab}K_{b t}=0
\ee
and
\be
\frac{1}{3}\gamma^{t}_{t}(-K^{3}+3KK_{2}-2 K_{3})=\frac{1}{3}\gamma^{t}_{t}(-\K^{3}+3\,\,\K\K_{2}-2\,\,\K_{3})\,.
\ee
So finally,
\be
T^{t}_{t}=\frac{1}{\lp^{3}}\big(- \K-\lambda L^{2}\big[(\R\K-2\, \R_{ij}\K^{ij})+\frac{1}{3}(-\K^{3}+3\,\K\K_{2}-2\,\K_{3})\big]\big)\,.\\
\ee

\section{EE for cylinder using Euclidean method}We will proceed here just like the sphere case and demonstrate the fact that $S_{JM}$ arises as a counterterm to remove all the power law divergences from the Euclidean action.
$$f(z)=z_{0}\big(1-\frac{1}{4 z_{0}^{2}} z{}^2+f_{4} z^4+\tilde f_4 z^4\ln z\big)\,.$$
  We will quote the results for the different parts of the action after expanding around small  $z$\,. 
\begin{align}
\begin{split}
I_{bulk}&=\frac{4 \pi   \beta  H L^3 (f_{0}^{2}-{ \Lambda^{2}}) (1-6 f_{\infty} \lambda )}{f_{\infty}^{3/2} \lp^3 \epsilon^4}-\frac{2 \pi  \beta   H L^3 (1-6 f_{\infty} \lambda)}{f_{\infty}^{3/2} \lp^3 \epsilon^2}\\&-\frac{\pi   \beta  H L^3 (1-6 f_{\infty} \lambda ) \left(32 f_{4} f_{0}^3+32 \tilde f_{4} f_{0}^3 \ln z+1\right)}{4  f_{\infty}^{3/2} \lp^3 f_{0}^{2} }\ln(\frac{f_{0}}{\epsilon})+\cdots\,.
\end{split}
\end{align}
\begin{align}
\begin{split}
\tilde I_{GH}&=-\frac{\pi  \beta  H L^3 (1-6 f_{\infty} \lambda )}{f_{\infty}^{3/2} \lp^3 \epsilon^2}+\frac{\pi  \beta  H L^3 \left(4 f_{0}^3 (4 f_{4}-3 \tilde f_{4}+4\tilde f_{4}\ln z) (1-6 f_{\infty} \lambda )+(1-4 f_{\infty} \lambda )\right)}{2 f_{\infty}^{3/2} \lp^3 f_{0}^{2} }\ln(\frac{f_{0}}{\epsilon})+\cdots\,.\\
\end{split}
\end{align}
Now,
\be
I_{GH}=\frac{4 \pi  \beta  H L^3 (f_{0}^{2}-{ \Lambda^{2}})  (1-2 f_{\infty} \lambda )}{f_{\infty}^{3/2} \lp^3 \epsilon^4}+\cdots\\
\ee
and,
\be
I_{ct}=-\frac{\pi   \beta  H L^3 (f_{0}^{2}-{ \Lambda^{2}})   (3-2 f_{\infty} \lambda )}{f_{\infty}^{3/2}\lp^3 \epsilon^4}+\cdots\\
\ee
And the last piece,
\begin{align}
\begin{split}
I_{JM}&=\frac{\beta}{4\pi } S_{JM}\,.\\
&=-\frac{\pi  \beta  H L^3 (1-6 f_{\infty} \lambda)}{2 f_{\infty}^{3/2} \lp^3 \epsilon ^2}+\frac{\pi  \beta  H L^3 (1-2 f_{\infty} \lambda )}{8 f_{\infty}^{3/2} \lp^3 f_{0}^{2}} \ln (\frac{f_{0}}{\epsilon} )+\cdots.
\end{split}
\end{align}
Adding them together we get,
\begin{align}
\begin{split}
I_{tot}&=\frac{3 \pi \beta   H L^3  \left(1-2 f_{\infty} \lambda -16 \tilde f_{4}f_{0}^3 (1-6 f_{\infty} \lambda )\right)}{8 f_{\infty}^{3/2} \lp^3 f_{0}^{2}}\ln (\frac{f_{0}}{\epsilon} )+\cdots.
\end{split}
\end{align}
Now when we solved for $f(z)$ around $z=0$, the $\ln$ term came as $\tilde f_{4} z^4 \ln z$. But here $z$  is
dimensionful so $\ln z$ should actually be $ \ln (z/{\Lambda_{scale}})$ where $\Lambda_{scale}$ is some arbitrary scale. At this level, we cannot determine what this scale is. So we do not have to set it to $\epsilon$. On the other hand, the $1/z$ terms that lead to $ (1-2 \lambda f_{\infty})\ln z$  must be replaced with $ (1-2 \lambda f_{\infty}) \ln( z/{\epsilon})$ where  the $\epsilon$ comes from the lower limit of the integral. Thus the $\tilde f_{4} \ln (z/{\Lambda_{scale}})$ and these terms can be distinguished. Now since this is a conformal field theory, it cannot matter what we choose the  scale to be. So it must be possible to remove these terms by choosing some scheme. One convenient way to choose a scheme is to set $\Lambda_{ scale}=\epsilon$ so that
the $ \tilde f_{4}$ terms are zero. In this scheme, 
\be
I_{tot}=\frac{3 \beta H c}{8\pi f_{0}^{2}}\ln (\frac{f_{0}}{\epsilon} )+\cdots
\ee
Note that there is no power law divergences in $I_{tot}$. As before, $T=\frac{\zeta}{f_{0}}$ where $\zeta$ is a $\mathcal{O}(1)$ number yet to be determined and $ \beta=\frac{1}{T}$. Unlike the sphere case $I_{tot}$ is dependent on the temperature.
\be
F=I_{tot}T=\frac{3  H c T^{2}}{8\pi \zeta^{2}}\ln (\frac{f_{0}}{\epsilon} )+\cdots\,.
\ee
Then,
\begin{align}
\begin{split}
S_{EE}=-\frac{\partial F}{\partial T}&= -\frac{3  H c T}{4\pi \zeta^{2}}\ln (\frac{f_{0}}{\epsilon} )+\cdots = -\frac{3  H c }{4\pi f_{0} \zeta}\ln (\frac{f_{0}}{\epsilon} )+\cdots\\
\end{split}
\end{align}
Identifying $\zeta= \frac{3}{2\pi}$ we get,
\be
S_{EE}= -\frac{H c }{2 f_{0} }\ln (\frac{f_{0}}{\epsilon} )+\cdots
\ee
which is the universal term for a cylinder. The key missing link in this computation is an independent reason for the choice of $\zeta$. In the sphere case note that $T$ worked out to be the same temperature in the map from the entanglement entropy of the sphere to the thermal entropy of a hyperboloid \cite{chm}.

\section{Details of the Calculation of Section 3}
We list here the curvature tensors required for the calculations in section 3 in an expansion in both $\epsilon$ and $w$.
\paragraph{Metric} $$ds^{2}=e^{2\rho}dw d\bar w+g_{ij}dx^{i}dx^{j}\,,$$
where $g_{ij}=h_{ij}+w\,^{(w)}\K_{ij}(x)+\bar w \,^{(\bar w)}\K_{ij}(x)+w\,\bar w\, Q_{ij}(x)$ and upto $\mathcal{O}(w^{2})$ the inverse metric is given by  \vspace{.3cm} \\ $g^{ij}=h^{ij}-w\,^{(w)}\K^{ij}-\bar w\,^{(w)}\K^{ij}+w^{2}\,\,^{(w)}\K^{i}_{k}\,^{(w)}\K^{kj}+\bar w^{2}\,\,^{(\bar w)}\K^{i}_{k}\,^{(\bar w)}\K^{kj}+w\,\bar w\, (2\,^{(w)}\K^{i}_{k}\,^{(\bar w)}\K^{kj}-Q^{ij})\,.$     \par\vspace{.3cm} Also  we have used $^{(w)}\K^{ij}= h^{il}h^{kj}\,^{(w)}\K_{ij}$ in the expression of $g^{ij}\,.$  $$\,^{(w)}\tilde \K_{ij}= \,^{(w)}\K_{ij}+\bar w \,Q_{ij}\quad ^{(\bar w)}\tilde \K_{ij}=\,^{(\bar w)}\K_{ij}+w\, Q_{ij}\,.$$
From here on  all the indices are lowered and raised using $g_{ij}\,.$ The $\nabla$'s and $^{d}R$  here are constructed using  $g_{ij}\,.$ The curvature tensors in section 3 involve $\R$'s which are made of $h_{ij}$. These are obtained from the $^{d}R$'s listed below.
\paragraph{Christoffel symbols}
\begin{align}
\begin{split}
{\Gamma^w}_{ww}&=2 \partial_w \rho\,,\quad{\Gamma^{\bar w}}_{\bar w\bar w}=2 \partial_{\bar w} \rho\,, \quad{\Gamma^w}_{ij}=-e^{-2 \rho}\,\,{}{} ^{(\bar w)}\tilde{\K}_{ij}\,,\quad  {\Gamma^{\bar w}}_{ij}=-e^{-2 \rho}\,\, ^{(w)} \tilde\K_{ij}\,,\\
{\Gamma^i}_{wj}&=\frac{1}{2}\,^{(w)}\tilde\K^{i}_{j}\,\,\,,\,\, {\Gamma^i}_{\bar wj}=\frac{1}{2}\,^{(\bar w)}\tilde\K^{i}_{j}\,\,\,,\,
\Gamma^{i}_{jk}=\frac{1}{2}g^{il}(\partial_{j}g_{lk}+\partial_{k}g_{lj}-\partial_{l}g_{jk})\,,\\
{\Gamma^{\bar w}}_{ww}&= {\Gamma^{ w}}_{\bar w\bar w}=\Gamma^{w}_{\bar w w}={\Gamma^ {w}}_{w j}={\Gamma^ {w}}_{\bar w j}={\Gamma^j}_{ww}={\Gamma^j}_{\bar ww}=0\,.\\
\end{split}
\end{align}

\paragraph{Riemann tensors}
 
\begin{align}
\begin{split}
{\hat R_{wiw}}{}^j &=\frac{1}{4}\,\,^{(w)}\tilde\K^{k}_i\,\,^{(w)}\tilde\K^{j}_k+(\partial_w \rho)\,\,^{(w)}\tilde\K^{j}_i-\frac{1}{2}g^{jk}(\partial_w\,^{(w)}\tilde\K_{ik})\,,\\ {\hat R_{ijw}}{}^w &=\frac{1}{2}e^{-2\rho}\,\,^{(\bar w)}\tilde\K_{ki}\,
\,^{(w)}\tilde\K^{k}_j-\frac{1}{2}e^{-2\rho}\,\,^{(\bar w)}\tilde\K_{kj}\,\,^{(w)}\tilde\K^{k}_i\,,\\
{\hat R_{wi\bar w}}{}^j &=\frac{1}{4}\,\,^{(w)}\tilde\K_{ki} \,\,^{(\bar w)}\tilde\K^{kj}-\frac{1}{2}g^{jk}(\partial_w \,^{(\bar w)}\tilde\K_{ki})\,,\quad {\hat R_{ijk}}^{w}=e^{-2\rho}\big(\nabla_i\,(^{(w)}\tilde\K_{jk})-\nabla_j\,(^{(\bar w)}\tilde\K_{ik})\big)\,,\\
{\hat R_{ijk}}{}^l &=-\frac{1}{2}e^{-2\rho}\,\,^{(\bar w)}\tilde\K_{ik}\,\,^{(\bar w)}\tilde\K^{ l}_j-\frac{1}{2}e^{-2\rho}\,\,^{(w)}\tilde\K_{ik}\,\,^{(w)}\tilde\K^{ l}_j+\frac{1}{2}e^{-2\rho}\,\,^{(w)}\tilde\K_{jk}\,\,^{(\bar w)}\tilde\K^{ l}_i+\frac{1}{2}e^{-2\rho}\,\,^{(\bar w)}\tilde\K_{jk}\,\,^{(w)}\tilde\K^{ l}_i \\ &~~~~~~+\,^d R_{ijk}{}^{l}\,,\\
{\hat R_{www}}{}^w&={\hat R_{ww\bar w}}{}^w={\hat R_{www}}{}^i={\hat R_{ww\bar w}}{}^i=0\,.\\
\end{split}
\end{align}
The components ${\hat R_{\bar w i \bar w}}{}^j$ etc can be read off easily from above. 


\paragraph{Ricci tensors}

\begin{align}
\begin{split}
\hat R_{ww}&=\frac{1}{4}\,\,^{(w)}\tilde\K_{ij}\,\,^{(w)}\tilde\K^{ij} +(\partial_w \rho)\,\,^{(w)}\tilde\K-\frac{1}{2}g^{ij}(\partial_w\,\,^{(w)}\tilde\K_{ij})\,,\,
\hat R_{\bar ww}=\frac{1}{4}\,\,^{(w)}\K_{ij} \,\,^{(\bar w)}\K^{ij}-\frac{1}{2}g^{ik}(\partial_w\,\,^{(\bar w)}\tilde\K_{ik})\,,\,\\
\hat R_{ij}&=e^{-2\rho}\big(\,\,^{(\bar w)}\tilde\K^{k}_i\,\,^{(w)}\tilde\K_{kj}+\,\,^{(\bar w)}\tilde\K^{k}_j\,\,^{(w)}\tilde\K_{ik}-\frac{1}{2}\,\,^{(\bar w)}\tilde\K_{ij}\,\,^{(w)}\tilde\K-\frac{1}{2}\,\,^{ (w)}\tilde\K_{ij}\,\,^{(\bar w)}\tilde\K -g^{l}_{j}(\partial_w \,^{(\bar w)}\tilde\K_{il})\\&~~~~~- g^l_j(\partial_{\bar w} \,^{(w)}\tilde\K_{il})\big)+^d\!R_{ij}\,,\\ \hat R_{iw}&=\frac{1}{2}\big(\nabla_j (^{(\bar w)}\tilde\K^j_{i})-\nabla_i (^{(w)}\tilde\K)\big)\,.\\
\end{split}
\end{align}

\paragraph{Ricci Scalar}
\be
\hat R=3e^{-2\rho}\,\,^{(w)}\tilde\K^{ij}\,\,^{(\bar w)}\tilde\K_{ij}-e^{-2\rho}\,\,^{(w)}\tilde\K\,\,^{(\bar w)}\tilde\K -2e^{-2\rho} g^{ik}(\partial_w \,^{(\bar w)}\tilde\K_{ik})-2 e^{-2\rho} g^{ik}(\partial_{\bar w} \,^{( w)}\tilde\K_{ik}) +^d\!\! R\,.
\ee

\end{document}